\ifx\pdfsuppresswarningpagegroup\undefined\else\pdfsuppresswarningpagegroup=1\fi\relax
\RequirePackage{silence}
\PassOptionsToPackage{unicode, pdfprintscaling=None, colorlinks}{hyperref}
\expandafter\relax\csname documentclass\endcsname[aps,prx,reprint,nofootinbib,floatfix,unsortedaddress,superscriptaddress,longbibliography,preprintnumbers]{revtex4-2}

\usepackage[normalem]{ulem}
\usepackage{orcidlink,multirow,siunitx,colortbl,dcolumn,ulem,tikz,graphicx,latexsym,amsmath,amssymb,amsfonts,bm,bbm,microtype,xspace,placeins,float,mathtools,xcolor,makecell,tabularx, booktabs, longtable, supertabular}

\usepackage[export]{adjustbox}
\usepackage{hyperref}
\usepackage[capitalise]{cleveref}
\usepackage{array}

\usetikzlibrary{decorations.markings, arrows, decorations.pathmorphing}
\graphicspath{{pic/}}
\allowdisplaybreaks

\newlength{\FigureWidth}
\newlength{\RowHeight}
\newcolumntype{L}[1]{>{\raggedright\arraybackslash\rule{0pt}{\RowHeight}}p{#1}}
\newcolumntype{C}[1]{>{\centering\arraybackslash\rule{0pt}{\RowHeight}}p{#1}}
\newcolumntype{R}[1]{>{\raggedleft\arraybackslash\rule{0pt}{\RowHeight}}p{#1}}
\newcolumntype{M}[1]{>{\centering\arraybackslash\rule{0pt}{\RowHeight}}m{#1}}
\newcommand\TopRule{\Xhline{0.08em}}

\newcommand\MidRule{\Xhline{0.03em}}
\newcommand\BotRule{\Xhline{0.08em}}

\makeatletter
\newcommand\showtitleinbib{{\escapechar=`\\ \immediate\write\@auxout{%
\csname citation{REVTEX42Control}\endcsname^^J%
\csname citation{apsrev42Control}\endcsname
}}}
\makeatother

\newcommand{\cH}{{\cal H}}

\newcommand{\chUp}{{\hat{\cal U}_P}}

\newcommand{\phys}{{\mathrm{phys}}}

\newcommand{\ldim}{{{\cal D}(\cH_s^g)}}

\newcommand{\ungroup}[1]{#1}
\newcommand{\withbreak}[1]{\expandafter\ungroup#1}

\def\dounbracket[#1]{#1}
\newcommand{\txtw}{{\mathsf{w}}}

\newcommand{\bsigma}{{\overline{\sigma}}}

\newcommand{\SU}{{\mathrm{SU}}}

\def\ket#1{\left| #1\right\rangle}

\renewcommand\vec\mathbf

\newcommand\redsout{\bgroup\markoverwith{\textcolor{red}{\rule[0.5ex]{2pt}{1.4pt}}}\ULon}

\newcommand\dootimesall[2]{\ifx0#1\else\mathbf{#1}\ifx0#2\else\def\mytmp{\otimes\dootimesall{#2}}\expandafter\expandafter\expandafter\mytmp\fi\fi}

\newcommand\bket[2]{\ket{\mathbf#1\mathbf#2}}

\newcommand{\RN}[1]{%
  \textup{\uppercase\expandafter{\romannumeral#1}}%
}
\usepackage{relsize}

\usepackage[acronyms, nohypertypes={acronym}, nopostdot, style=super, nonumberlist, toc]{glossaries}
\setacronymstyle{long-sc-short}
\glsenableentrycount

\newcommand\ac[1]{\gls{#1}}

\newcommand\acp[1]{\glspl{#1}}

\newacronym{PNA}{pna}{particle-number-algorithm}
\newacronym{SFA}{sfa}{spin-flip-algorithm}

\newacronym{WF}{wf}{Wilson-Fisher}
\newacronym{AF}{af}{asymptotically free}

\newacronym{RG}{rg}{renormalization group}

\newacronym{QIS}{qis}{Quantum Information Science}
\newacronym{PPT}{ppt}{positive-semidefinite partial transpose}
\newacronym{KS}{ks}{Kogut-Susskind}
\newacronym{NPT}{npt}{negative partial transpose}

\newacronym{AS}{as}{Anti-Symmetric}

\newacronym[longplural={conformal field theories}]{CFT}{cft}{conformal field theory}
\newacronym[longplural={lattice field theories}]{LFT}{lft}{lattice field theory}
\newacronym[longplural={effective field theories}]{EFT}{eft}{effective field theory}
\newacronym[longplural={quantum field theories}]{QFT}{qft}{quantum field theory}
\newacronym[longplural={lattice gauge theories}]{LGT}{lgt}{lattice gauge theory}
\newacronym[longplural={monomer-dimer tensor-networks}]{MDTN}{mdtn}{monomer-dimer tensor-network}

\newacronym{YM}{ym}{Yang-Mills}

\newacronym[]{DMRG}{dmrg}{Density Matrix Renormalization Group}
\newacronym[]{TFIM}{tfim}{Transverse Field Ising Model}
\newacronym[]{ICFT}{icft}{Ising-{\acrshort{CFT}}}
\newacronym[]{E8QFT}{e8qft}{$E_8$-{\acrshort{QFT}}}

\newacronym[]{LOCC}{locc}{Local Operations and Classical Communicaton}
\newacronym[]{OBC}{obc}{open boundary conditions}

\newacronym{MPS}{mps}{matrix product states}

\newacronym{JLP}{jlp}{Jordan-Lee-Preskill}

\newacronym{BBN}{bbn}{big bang nucleosynthesis}

\newacronym{LEC}{lec}{low-energy constant}

\newacronym{QCD}{qcd}{quantum chromodynamics}
\newacronym{MC}{mc}{Monte Carlo}

\newacronym{IR}{ir}{infrared}
\newacronym{UV}{uv}{ultraviolet}

\newacronym{QED}{qed}{quantum electrodynamics}
\newacronym{SNR}{snr}{signal-to-noise ratio}

\newacronym{NLSM}{nlsm}{nonlinear sigma model}

\newacronym{CL}{cl}{Complex Langevin}

\newacronym{CSA}{csa}{Cartan subalgebra}

\newacronym{SSB}{ssb}{spontaneous symmetry breaking}

\newacronym{AFQMC}{afqmc}{auxiliary field quantum Monte Carlo}
\newacronym{iHMC}{ihmc}{imaginary-mass Hybrid Monte Carlo}

\newacronym{MCMC}{mcmc}{Markov Chain Monte Carlo}

\newacronym{QI}{qi}{quantum information}

\newacronym{irrep}{{\rm irrep}}{unitary irreducible representation}

\newacronym{ASQR}{asqr}{antisymmetric qubit regularization}

\begin{document}

\title{Asymptotic-freedom and massive glueballs in a qubit-regularized SU(2) gauge theory}
\author{Rui Xian Siew\,\orcidlink{0000-0002-0745-8853}}%
 \email{ruixian.siew@duke.edu}
\affiliation{ Department of Physics, Box 90305, Duke University, Durham, North Carolina 27708, USA}
\author{Shailesh Chandrasekharan\,\orcidlink{0000-0002-3711-4998}}
\email{sch27@duke.edu}
\affiliation{ Department of Physics, Box 90305, Duke University, Durham, North Carolina 27708, USA}
\author{Tanmoy Bhattacharya\,\orcidlink{0000-0002-1060-652X}}
\email{tanmoy@lanl.gov}
\affiliation{Theoretical Division, Los Alamos National Laboratory, Los Alamos, New Mexico 87545, USA}

\date{\today}

\begin{abstract}
We argue that a simple qubit-regularized $\SU(2)$ \ac{LGT} on a plaquette chain serves as a pseudo-one-dimensional toy model for \ac{YM} theory in three spatial dimensions. We map the chain Hamiltonian to the \ac{TFIM} in a uniform magnetic field and demonstrate that it can be tuned to a continuum limit in which the short-distance physics is governed by the asymptotically free Ising \ac{CFT} describing free Majorana fermions, while the long-distance regime contains massive excitations of the $E_8$ \ac{QFT} that can be interpreted as one-dimensional analogues of glueballs. Furthermore, we find $\sqrt{\sigma}/m_1 = 0.249(1)$ where $\sigma$ is the string tension between two static quarks and $m_1$ is the mass of the lightest glueball.
\end{abstract}

\preprint{LA-UR-25-31739}

\maketitle

\section{Introduction}
\label{sec1}

Qubit regularization of \acp{QFT} is a framework for recovering a continuum \ac{QFT} as the limit of low-energy sectors of increasingly large spatial lattices of quantum mechanical systems with a fixed local Hilbert space~\cite{Chandrasekharan:2025Cb}, often describable as a qubit-model. If successful, this approach can provide a way to perform efficient \ac{QFT} calculations on a quantum computer \cite{Bauer:2023qgm}.
The success of this approach is deeply connected to the ideas of universality and renormalization, as originally introduced by Wilson~\cite{RevModPhys.55.583}. In particular, it is essential that the qubit-regularized theory possesses a critical point such that, in its vicinity, the long-distance physics is governed by the \ac{RG} fixed point corresponding to the desired continuum \ac{QFT}. This behavior is not guaranteed in general, and an important direction of research is to characterize the critical points of various qubit-regularized theories and to uncover the physics of the fixed points that emerge near them.

\begin{figure}[t]
\centering
\includegraphics[width=0.45\textwidth]{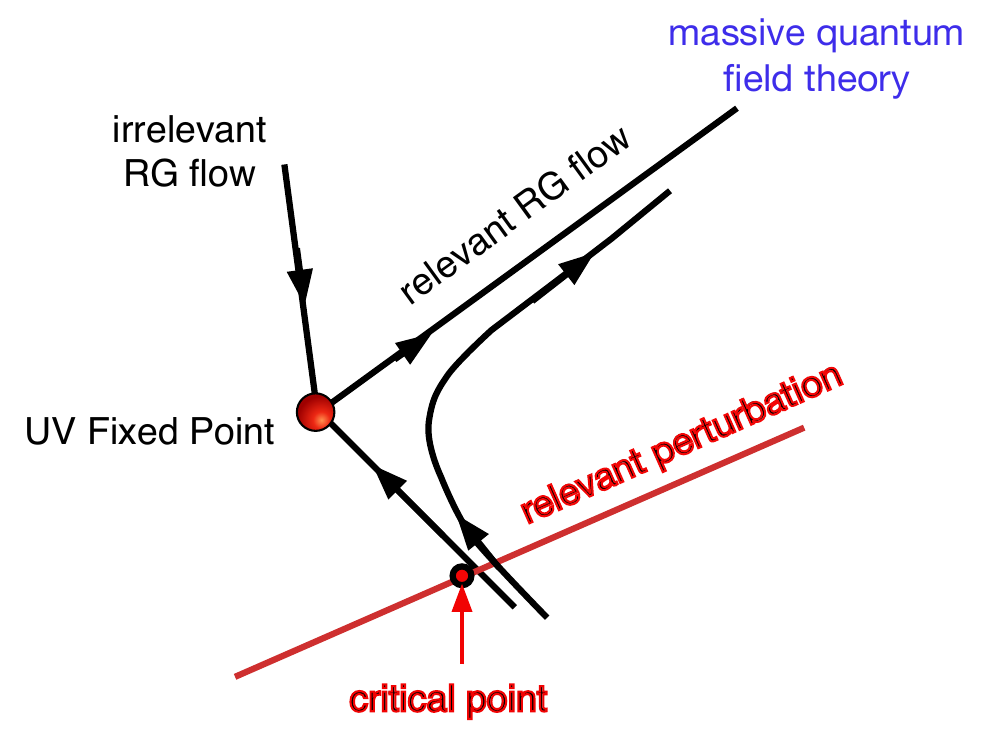}
\caption{A schematic illustration of the traditional \ac{RG} flows in the space of lattice models that reproduce a massive continuum \ac{QFT} in the \ac{IR} limit. Typically, one begins with a lattice model tuned to a critical point and then adds a relevant perturbation. The short-distance physics of the resulting massive \ac{QFT} is governed by the \ac{UV} fixed point, which can also be viewed as the \ac{IR} fixed point of the unperturbed lattice model at criticality.}
\label{fig:RGflow-trad}
\end{figure}

Massive \acp{QFT} of interest in particle physics typically arise through relevant perturbations of such \ac{RG} fixed points. These fixed points are characterized as \ac{UV} fixed points, and the associated scale-invariant---often conformally invariant---theories describe the short-distance, or equivalently high-energy, physics of the corresponding massive \ac{QFT}. Asymptotic freedom of pure non-Abelian gauge theories is a prime example of this feature. 
A major challenge for the Wilson lattice regularization program is to construct lattice models that exhibit this behavior: i.e., they have a critical point whose long-distance physics is governed by the desired \ac{UV} fixed point. When a small relevant perturbation is added, the continuum \ac{UV} physics manifests as an intermediate crossover regime, while the extreme long-distance behavior reproduces the desired massive \ac{QFT}. A schematic illustration of this traditional \ac{RG} flow in the space of lattice models is shown in \cref{fig:RGflow-trad}. 

In this work, we present an explicit example of such a traditional \ac{RG} flow realized in a qubit-regularized $\SU(2)$ \ac{LGT}. In our construction, the \ac{UV} fixed point is described by the two-dimensional \ac{ICFT}, while the corresponding massive theory is the \ac{E8QFT} of Zamolodchikov~\cite{Zamolodchikov:1989fp}.

The qubit-regularized $\SU(2)$ \ac{LGT} considered here is formulated in the Hamiltonian framework on a spatial lattice with the geometry of a plaquette chain. Such a chain is pseudo-one-dimensional in the sense that it is one-dimensional in the thermodynamic limit, but has local gauge `plaquette' excitations. In particular, the physical Hilbert space has scalar adjoint `gluons' at each lattice site that are the remnant of the single transverse gluon degree of freedom per site in the theory where the ladder is extended to form a genuinely two-dimensional lattice. This view point can be extended to $2+1$ dimensions where glue balls arise as weakly coupled system of heavy adjoint particles \cite{PhysRevD.75.101702}.

Hamiltonian \acp{LGT} on plaquette chains have been explored recently in several contexts~\cite{Nyhegn:2020cxu,Pradhan:2022lzo,Yao:2023pht,PhysRevD.110.014505,PhysRevD.111.094502,Chandrasekharan:2025smw}. Here, we extend these investigations by studying relevant perturbations away from critical points where massive continuum \acp{QFT} emerge in the \ac{IR}. In particular, we identify one such critical point in a specific $\SU(2)$ gauge theory on a plaquette chain by mapping the chain Hamiltonian to the \ac{TFIM} in a uniform magnetic field, 
\begin{align}
H \ =\ \sum_i \big(-\sigma_i^z \sigma_{i+1}^z - \sigma_i^x - h \sigma_i^z \big)\,.
\label{eq:TFIM}
\end{align}
It is well known that this Hamiltonian is critical at $h=0$, where it flows to the \ac{ICFT}. For small nonzero $h$, the long-distance physics is instead described by a massive \ac{QFT} with $E_8$ symmetry, as predicted by Zamolodchikov~\cite{Zamolodchikov:1989fp}. In a qubit-regularized $\SU(2)$ \ac{LGT}, we interpret the massive states as one-dimensional analogues of glueballs. Using the standard step-scaling function, we demonstrate that the continuum theory’s short-distance behavior is governed by the \ac{ICFT}, consistent with the traditional \ac{RG} flow shown in \cref{fig:RGflow-trad}. Because the \ac{ICFT} describes free Majorana fermions, we argue that the $\SU(2)$ plaquette chain also exhibits asymptotic freedom, much like a three-dimensional \ac{YM} theory.

\section{The SU(2) plaquette Chain}
\label{sec2}

We recently showed that the physical Hilbert space of $\SU(N)$ Hamiltonian \acp{LGT} can be constructed efficiently using the \ac{MDTN} basis $|\{\lambda_s\},\{\lambda_\ell\},\{\alpha_s\}\rangle$~\cite{Chandrasekharan:2025smw}. Each \ac{MDTN} basis state is characterized by three sets of labels defined as follows: (i) $\{\lambda_s\}$ denotes the configuration of \acp{irrep} of $\SU(N)$ associated with matter fields on the lattice sites, (ii) $\{\lambda_\ell\}$ denotes the configuration of \acp{irrep} associated with gauge fields on the lattice links, and (iii) $\{\alpha_s\}=1,2,\ldots,\ldim$ labels the degeneracy of singlet representations in the local site Hilbert space $\cH_s^g$, which is the tensor product of the \acp{irrep} of the links $\{\lambda_{\ell_s}\}$ connected to the site $s$ and the site representation $\lambda_s$.

\begin{figure}[b]
\centering
\includegraphics[width=0.45\textwidth]{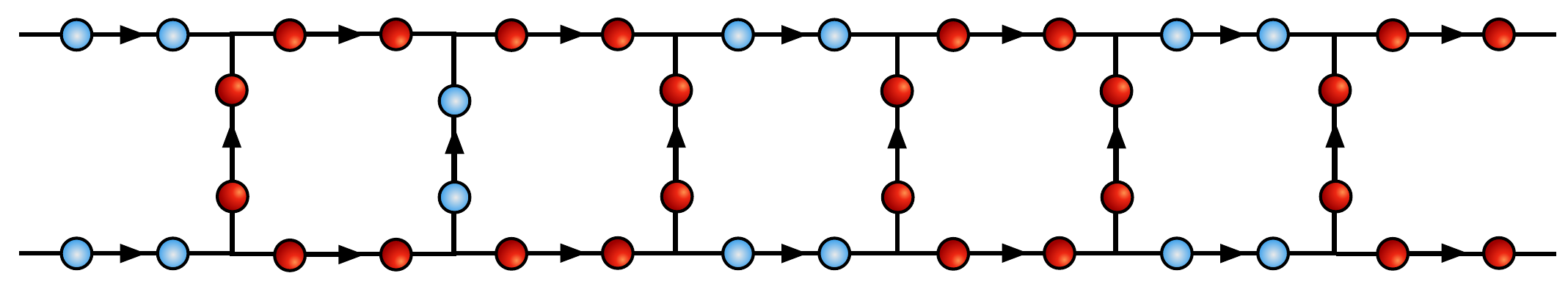}
\caption{A pictorial representation of an MDTN basis state in the plaquette chain with $\SU(2)$ gauge fields. The blue dots represent $\lambda = \mathbf{1}$ (singlets) and the red dots represent $\lambda = \mathbf{2}$ (doublets).}
\label{fig:pchain}
\end{figure}

\begin{figure}[t]
\centering
\includegraphics[width=0.35\textwidth]{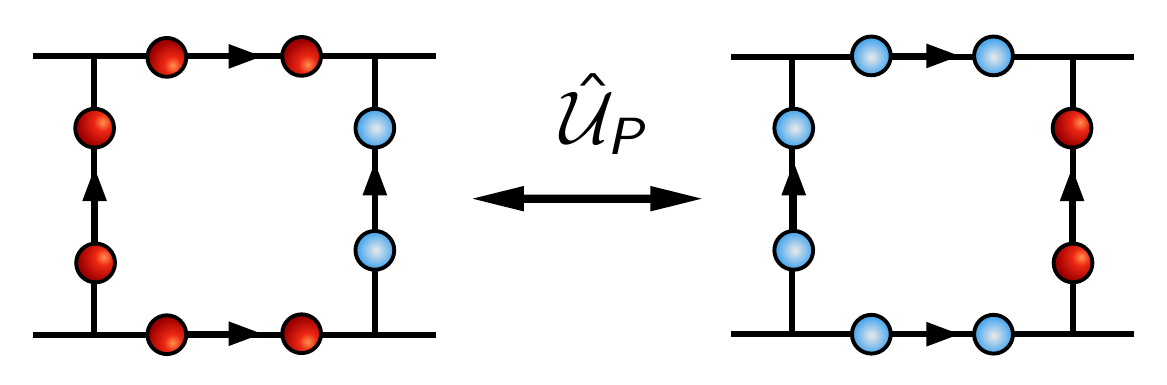}
\caption{A pictorial representation of the action of the plaquette operator $\hat{{\cal U}}_P$.}
\label{fig:Up}
\end{figure}

In this work, we focus on the $\SU(2)$ pure gauge theory, which contains no matter fields, allowing us to set all $\lambda_s = \mathbf{1}$ (singlets). In the conventional $\SU(2)$ \ac{LGT}, the link representations $\lambda_\ell$ can take any \ac{irrep} of $\SU(2)$ from the set $\{\mathbf{1}\ \text{(singlet)},\ \mathbf{2}\ \text{(doublet)},\ \mathbf{3}\ \text{(triplet)},\ \ldots\}$. In contrast, qubit regularization restricts these to a finite subset. The simplest such scheme involves only $\lambda_\ell = \{\mathbf{1}, \mathbf{2}\}$~\cite{Liu:2021tef}. A natural question then arises: can we construct qubit-regularized $\SU(2)$ \acp{LGT} within this minimal framework that exhibit nontrivial quantum critical points? Even more intriguing is the possibility of realizing massive continuum \acp{QFT} in the infrared by introducing relevant perturbations at these critical points. Recent studies have demonstrated that such behavior can indeed occur in qubit-regularized lattice spin models~\cite{Bhattacharya:2020gpm,Maiti:2023kpn}. Here, we show that analogous behavior emerges in the qubit-regularized $\SU(2)$ plaquette chain.

The plaquette chain has the geometry of a ladder composed of two one-dimensional chains of lattice sites connected by rungs that link neighboring sites across the chains. We consider a system containing $L$ plaquettes, which are connected periodically across the boundary. As stated, in the simplest $\SU(2)$ qubit regularization discussed above, the Hilbert space on each link contains two types of dimer tensors, $\mathbf{0}$ and $\mathbf{2}$. A pictorial representation of an \ac{MDTN} basis state for the plaquette chain is shown in \cref{fig:pchain}.

The ladder geometry of the plaquette chain allows us to distinguish between two types of gauge links: those along the chains, which we label as $\ell_c$, and those on the rungs connecting the two chains, labeled as $\ell_r$. In addition, we define plaquettes $P$ as the elementary squares formed by four such links. Using this notation, a generic gauge-invariant Hamiltonian for the plaquette chain can be written as
\begin{align}
H \ =\ \kappa_c \sum_{\ell_c}  
\hat{{\cal E}}_{\ell_c} + \kappa_r \sum_{\ell_r}  
\hat{{\cal E}}_{\ell_r} - \delta \sum_{P} \hat{{\cal U}}_P,
\label{eq:Hchain}
\end{align}
where the operator $\hat{{\cal E}}_\ell = (1 - \delta_{\lambda, \mathbf{1}})$ is diagonal in the \ac{MDTN} basis and assigns distinct energies to the different \acp{irrep} on each link $\ell$. In contrast, the plaquette operator $\hat{{\cal U}}_P$ is off-diagonal in the \ac{MDTN} basis and flips all the \acp{irrep} on the four links forming the plaquette, as illustrated in \cref{fig:Up}.

As explained in~\cite{Chandrasekharan:2025smw}, in the absence of matter fields the $\mathbf{2}$ links form closed loops. These loops can either wind around the boundary or lie entirely within a finite region of the ladder. Consequently, the physical Hilbert space splits into a direct sum of two topologically distinct sectors, depending on the number of loops that cross the boundary: $\cH_\phys^E$ (even sector, where zero or two loops cross the boundary) and $\cH_\phys^O$ (odd sector, where exactly one loop crosses the boundary). The Hamiltonian in \cref{eq:Hchain} does not couple these two sectors. Moreover, the two Hilbert-space sectors can be distinguished locally by examining the link \acp{irrep} on the top and bottom chains. Focusing on the two horizontal links of a given plaquette $P$, there are four possible basis states: $\bket11$, $\bket12$, $\bket21$, and $\bket22$, where the first label corresponds to the link \ac{irrep} on the top chain and the second to that on the bottom chain. In $\cH_\phys^E$, each plaquette can be in either $\bket11$ or $\bket22$, while in $\cH_\phys^O$ it can be in either $\bket12$ or $\bket21$. Thus, the local plaquette Hilbert space $\cH_{\cal P}$ spanned by the basis states $\bket11$ and $\bket22$ or $\bket12$ and $\bket21$ determines the full Hilbert space. The former belongs to the $\cH_\phys^E$ space and the latter belongs to the $\cH_\phys^O$ space.

\begin{figure}[t]
\centering
\includegraphics[width=0.45\textwidth]{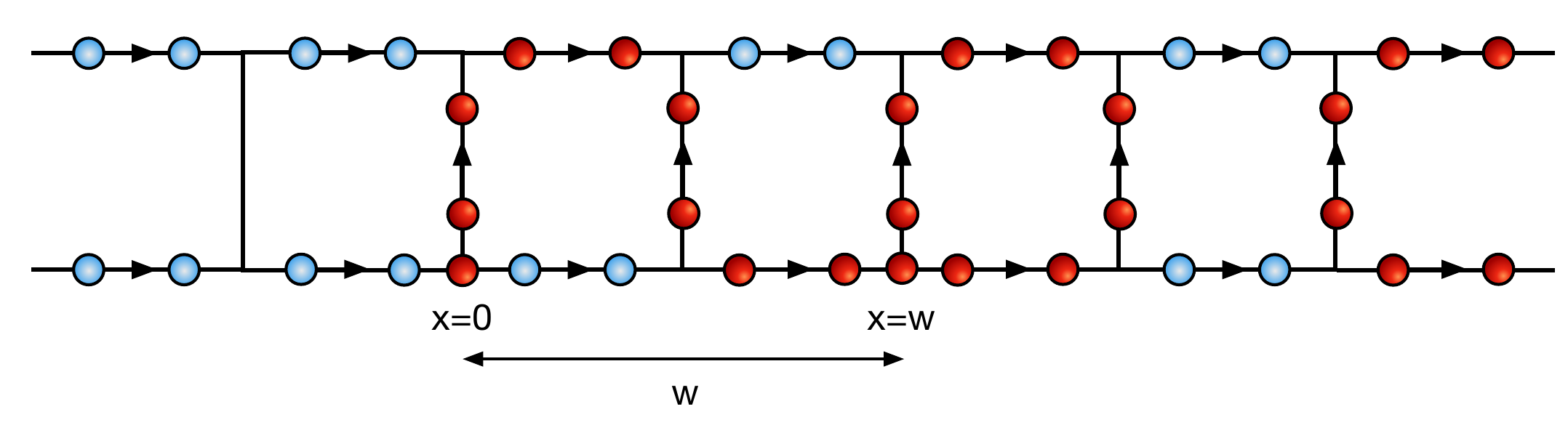}
\caption{A pictorial representation of an \ac{MDTN} basis state of the $\SU(2)$ plaquette chain with two heavy matter fields in the doublet representation located on the bottom rung at $x=0$ and $x=\txtw$.}
\label{fig:pchain+xy}
\end{figure}

The structure becomes richer when matter fields are introduced. The simplest case arises when two static matter fields in the doublet representation are placed at the bottom rungs of distinct lattice sites, say $x=0$ and $x=\txtw$. We view these as introducing heavy probe quarks in our pure gauge theory. In the presence of these heavy quarks, the local plaquette Hilbert space $\cH_{\cal P}$ of the gauge theory changes from $\cH_\phys^E$ to $\cH_\phys^O$ (or vice versa) at the sites $x=0$ and $x=\txtw$, depending on the configuration of the surrounding links. We can now distinguish two further Hilbert-space sectors, $\cH_\phys^\txtw$ and $\cH_\phys^{(L-\txtw)}$. In the former, the plaquette Hilbert spaces between the sites $x=0$ and $x=\txtw$ belong to the $\cH_\phys^O$ sector, while in the latter they belong to the $\cH_\phys^E$ sector. An illustration of an \ac{MDTN} basis state with two heavy matter fields in the $\cH_\phys^\txtw$ sector is shown in \cref{fig:pchain+xy}. 

When the sites $x=0$ or $x=\txtw$ contain four doublets (one from the heavy quark and three from the links connected to the site), there are two distinct ways to satisfy Gauss's law locally. This freedom is encoded in the index $\alpha_s$ in the \ac{MDTN} basis. In this situation it is straightforward to verify that the action of $\chUp$ on each of these two states produces the same unique state. Consequently, a particular linear combination of the two basis states is always annihilated by $\chUp$ and therefore remains inert under the dynamics generated by \cref{eq:Hchain}. Operationally, it is as if the operator $\chUp$ is effectively removed from the two adjacent plaquettes attached to the site $x=0$ or $x=\txtw$ whenever this special linear combination is chosen in the \ac{MDTN} basis.

Although one could in principle study the resulting effect of removing $\chUp$ at the locations of the heavy matter fields, in this work we choose, for simplicity, to work with the orthogonal linear combination that is not annihilated by $\chUp$. This ensures that $\chUp$ remains non-zero on all plaquettes, as in the original construction. 

Introducing additional heavy quarks allows us to probe the rich gauge-theory Hilbert space more completely. Fixing the locations of the quarks defines new sectors of the theory, and the physics described by the Hamiltonian in each sector can differ in nontrivial ways. As we discuss below, within each such sector, the Hamiltonian in \cref{eq:Hchain} can be mapped to a distinct \ac{TFIM}.

\section{Mapping to the Ising Model}
\label{sec3}

\begin{figure}[h]
\centering
\includegraphics[width=0.12\textwidth]{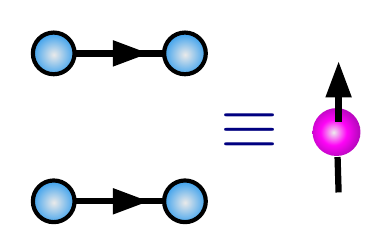}
\hspace{0.2in}
\includegraphics[width=0.12\textwidth]{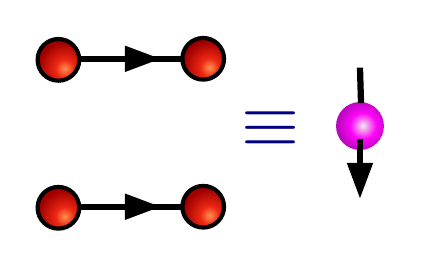}
\caption{Mapping the $\bket11$ to $\ket{\uparrow}$ and $\bket22$ to $\ket{\downarrow}$.}
\label{fig:HSmap}
\end{figure}

\begin{figure}[h]
\centering
\includegraphics[width=0.3\textwidth]{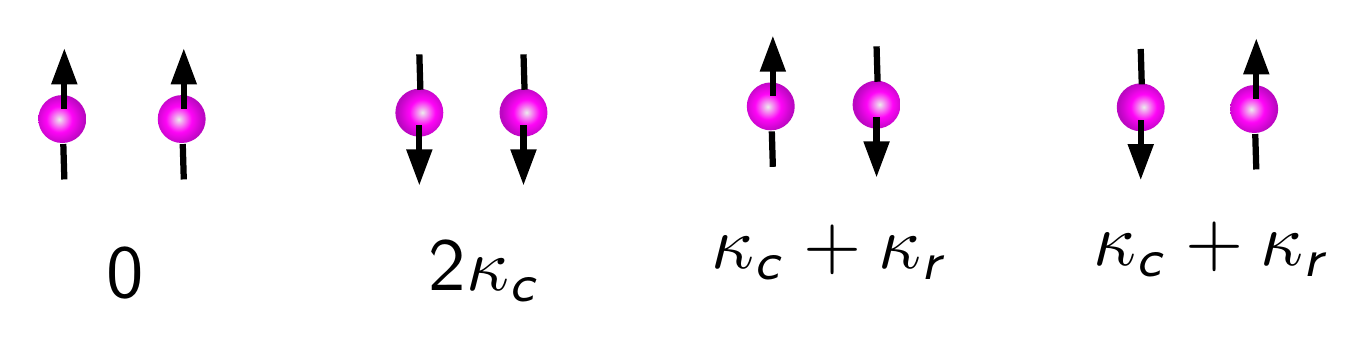}
\caption{Interaction energies of nearest neighbor spins. These energies are encoded in the diagonal matrix elements of \cref{eq:HIsing-E}.}
\label{fig:diagwts}
\end{figure}

Let us first focus on the physics of $H$ restricted to the even sector defined through the Hilbert space $\cH_\phys^E$. In this Hilbert space we can map the two possible plaquette states $\bket11$ and $\bket22$ to the quantum Ising Hilbert space of $\ket{\uparrow}$ and $\ket{\downarrow}$ respectively. This mapping is illustrated pictorially in \cref{fig:HSmap}. 
In this mapping the Hamiltonian of \cref{eq:Hchain}, in the $\cH_\phys^E$ sector, can be rewritten as
\begin{align}
H^E \ =\ 
\sum_i\ \Big(\kappa_c + \frac{\kappa_r}{2} - \frac{\kappa_r}{2} \sigma_i^z \sigma_{i+1}^z - \kappa_c \sigma_i^z - \delta \sigma_i^x\Big),
\label{eq:HIsing-E}
\end{align}
where $\sigma_i^z$ and $\sigma_i^x$ are Pauli matrices associated to the plaquette ${\cal P}$ labeled by $i$. The states $\ket{\uparrow}$ and $\ket{\downarrow}$, illustrated in \cref{fig:HSmap}, are eigenstates of $\sigma_i^z$ with eigenvalue $+1$ and $-1$ respectively. It is easy to see that the operator $\hat{{\cal U}}_P$ maps to $\sigma_i^x$. 
Similarly, the terms diagonal in this basis, along with the energies of the rung link are then given by terms in \cref{fig:diagwts}.
These energies then allow us to Hamiltonian in \cref{eq:HIsing-E}. 

\begin{table*}[t]
\centering
\renewcommand{\arraystretch}{1.4}
\setlength{\tabcolsep}{4pt}
\begin{tabular}{|c|c|c|c|c|c|c|c|c|}
\TopRule
$\kappa_c$ & $L$ & $\mathcal{E}_0(\txtw=0)$ & $m_1$ & $m_2$ & $\bsigma$ & $m_2/m_1$ & $\sqrt{\sigma}/m_1$ & $\kappa_c^{8/15}L$ \\
\MidRule
0.3 & 12 & -18.273(3) & 2.834(3) & 4.488(3) & 0.2511(5)& 1.584(3) &  0.2497(10) &  6.31 \\
0.3 & 16 & -24.363(3) & 2.834(3) & 4.548(3) & 0.2507(3) & 1.604(3) &  0.2496(09) & 8.42\\
0.2 & 16 & -22.966(3)  & 2.287(3) & 3.639(3) & 0.1631(3) & 1.591(3) &  0.2494(10) & 6.78\\
0.2 & 24 & -34.448(2) & 2.287(2) & 3.679(2) & 0.1628(1) & 1.609(2) &  0.2492(07)  & 10.17\\
0.1 & 28 & -37.821(3) & 1.583(3) & 2.555(3) & 0.0779(1)  & 1.614(5) &  0.2490(11) & 8.20 \\
0.1 & 32 & -43.224(3) & 1.583(3) & 2.555(3) & 0.0779(1) & 1.614(5) &  0.2490(10) & 9.37\\
0.1 & 48 & -64.836(2) & 1.584(2) & 2.557(2) & 0.0777(1) & 1.614(3) &  0.2486(9) & 14.06\\
0.01 & 96 & -122.8701(4) & 0.4649(4) & 0.7503(4) & 0.00671(1) & 1.614(2) &  0.2489(9) & 8.23 \\
\BotRule 
\end{tabular}
\caption{\label{tab:string_tension}
Using \ac{DMRG}, we extract the ground-state energy of the lattice Hamiltonian \cref{eq:HIsing-E} (after dropping the constant term) at $\kappa_r = 2$ and $\delta = 1$. This quantity is denoted by ${\cal E}_0(\txtw = 0)$. We also compute the two lowest glueball masses, $m_1$ and $m_2$, as well as the string tension $\bsigma$, obtained from a linear fit to \cref{eq:stringT}. The errors shown are determined from the fluctuations of the data points about the expected fit form, such that linear fits of the type shown in \cref{fig:stringtension_DMRG} have a $\chi^2/\text{d.o.f.}$ less than one. The string tension in units of energy-squared is given by $\sigma = \zeta\, \bsigma$ with $\zeta=1.995(7)$ as explained in the text. The dimensionless ratios $m_2/m_1$ and $\sqrt{\sigma}/m_1$ are expected to be universal quantities in the continuum gauge theory that depend on the scaling variable $\mu = \kappa_c^{8/15} L$, shown in the last column, in the limits $L \to \infty$ and $\kappa_c \to 0$. The \ac{E8QFT} emerges when $\mu$ becomes large.
}
\end{table*}

We can repeat the above analysis to map the physics of $H$ when restricted to the odd sector defined through the Hilbert space $\cH_\phys^O$. The final answer is 
can be rewritten as
\begin{align}
H^O \ =\ 
\sum_i\ \Big(\kappa_c + \frac{\kappa_r}{2} - \frac{\kappa_r}{2} \sigma_i^z \sigma_{i+1}^z - \delta \sigma_i^x\Big),
\label{eq:HIsing-O}
\end{align}
where $\sigma_i^z$ and $\sigma_i^x$ are Pauli matrices associated to the plaquette ${\cal P}$ labeled by $i$ as before, but now act on the $\bket12$ and $\bket21$ Hilbert space. 

We can extend this analysis to include heavy quarks as discussed in the previous section. Such  `static charges\rlap', allow us to compute the static quark potential, from which we can compute the string tension $\sigma$ in this linearly confining theory. In the next section we will argue that $\cH_\phys^E$ is a natural sector to explore the physics of plaquette chain. We can then compute the heavy quark potential by considering the ground state energies of the Hamiltonian of the theory in the Hilbert space sector $\cH_\phys^\txtw$ as a function of $\txtw$, assuming $\txtw < L/2$. The Hamiltonian of the theory in this sector is then given by
\begin{align}
H^\txtw \ =\ 
\sum_i\ \Big(\kappa_c + \frac{\kappa_r}{2} - \frac{\kappa_r}{2} \sigma_i^z \sigma_{i+1}^z - 
\kappa_c \theta_i \sigma_i^z - 
\delta \sigma_i^x\Big),
\label{eq:HIsing-w}
\end{align}
where $\theta_i = 0$ for plaquettes between the sites $x=0$ and $x=\txtw$ and $\theta_i = 1$ outside. Comparing \cref{eq:HIsing-w} and \cref{eq:HIsing-E} we note that when $\txtw=0$, $H^\txtw = H^E$.

\section{Massive Glueballs}
\label{sec4}

In a confining theory, loops of $\mathbf{2}$-links that wind around space are energetically suppressed, acquiring an energy proportional to the system size except at criticality. This observation suggests that the even sector, $\cH_\phys^E$, should be identified as the physical Hilbert space of local gauge-invariant excitations over the infinite-volume ground-state. In this sector, as argued in \cref{sec3}, the Hamiltonian of the qubit-regularized $\SU(2)$ gauge theory on the plaquette chain can be mapped to \cref{eq:HIsing-E}, which becomes identical to \cref{eq:TFIM} upon setting $\kappa_r = 2$, $\kappa_c = h$, and $\delta = 1$.

The physics of \cref{eq:TFIM} has been extensively studied, and it is well known that for small nonzero values of $h$ the model flows to a massive \ac{QFT} with hidden $E_8$ symmetry in the \ac{IR}~\cite{Zamolodchikov:1989fp,Delfino:1995zk,Delfino:1996jr}. Zamolodchikov’s analysis predicts eight stable particles in the continuum \ac{QFT}. However, except for the three lowest masses $m_1$, $m_2$, and $m_3$, the higher masses lie above the two-particle thresholds of lighter states. These lightest three masses can be extracted from the four lowest energy levels, ${\cal E}_i, i=0,1,2,3$, of \cref{eq:TFIM} at zero momentum using the relation $m_i = {\cal E}_i - {\cal E}_0$.  
In the limits $L \rightarrow \infty$ and $h \rightarrow 0$, these masses are expected to satisfy the analytic predictions~\cite{Zamolodchikov:1989fp},
\begin{align}
\frac{m_2}{m_1} = 2\cos\frac{\pi}{5} \approx 1.61803, \qquad
\frac{m_3}{m_1} = 2\cos\frac{\pi}{30} \approx 1.98904,
\label{eq:E8pred}
\end{align}
which have been confirmed in several studies~\cite{Fonseca:2006au,Jha:2024jan,Karna:2025pof}.

\begin{figure}[t]
\centering
\includegraphics[width=0.47\textwidth]{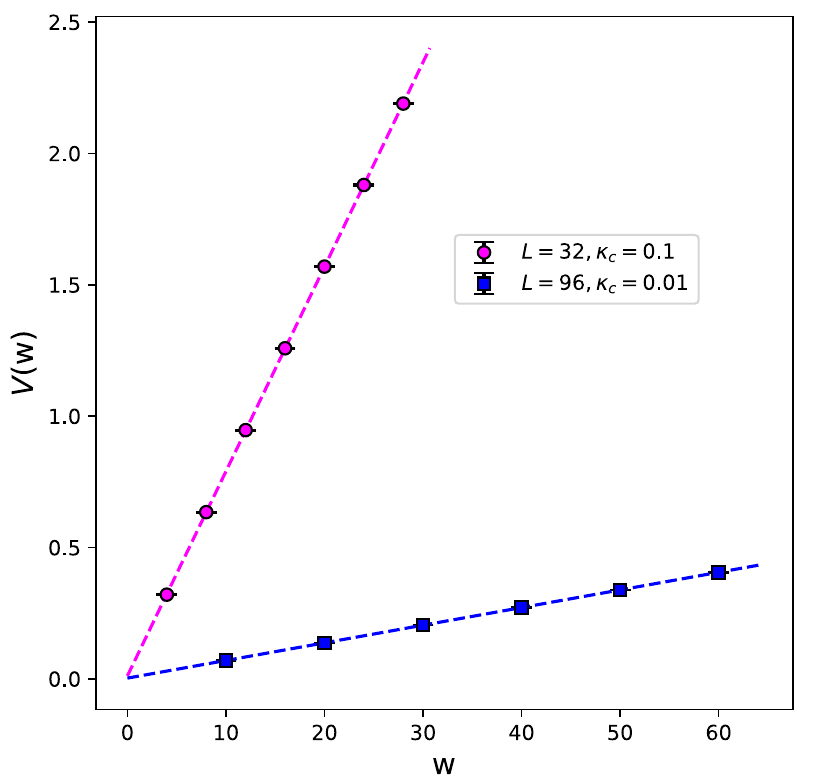}
\caption{Plot of the static quark potential $V(\txtw)$ as a function of $\txtw$ at $(L = 32, \kappa_c = 0.1)$ and $(L = 96, \kappa_c = 0.01)$. These calculations were performed using \ac{DMRG}. Because the uncertainties in the computed energies are difficult to estimate---and may be small enough that the data become sensitive to corrections beyond the linear form of \cref{eq:stringT}---we assigned errors of $0.002$ to the $L=32$ data and $0.0004$ to the $L=96$ data, consistent with the observed fluctuations about the fit; that is, these choices yield $\chi^2/d.o.f \sim 1$ for the linear fit.}
\label{fig:stringtension_DMRG}
\end{figure}

The above discussion can also be interpreted as being applicable to the qubit-regularized $\SU(2)$ gauge theory on the plaquette chain, restricted to the Hilbert space $\cH_\phys^E$ and defined by the Hamiltonian \cref{eq:Hchain} with $\kappa_r = 2$, $\delta = 1$, and $\kappa_c \rightarrow 0$, describing a continuum \ac{QFT} with massive particles that can be interpreted as pseudo-one-dimensional analogues of glueballs. The lowest three glueball masses satisfy the relations in \cref{eq:E8pred}.

We can also compute the static quark potential $V(\txtw)$ between two heavy quarks as a function of their separation $\txtw$ by evaluating the ground-state energy ${\cal E}_0(\txtw)$ of $H^\txtw$, defined in \cref{eq:HIsing-w}. By fixing $\kappa_r = 2$ and $\delta = 1$, taking $L$ to be sufficiently large, and choosing a variety of values of $\kappa_c$, confinement implies a functional dependence of the form
\begin{align}
V(\txtw)\ =\ {\cal E}_0(\txtw) - {\cal E}_0(0) \ =\ \alpha + \bsigma\, \txtw,
\label{eq:stringT}
\end{align}
in the regime $1 \ll \txtw \ll L$. The string tension $\bsigma$ is expected to depend strongly on $\kappa_c$ but, once $L$ is sufficiently large, should be independent of the system size.
We have verified these expectations by computing ${\cal E}_0(\txtw)$ using the \ac{DMRG} method and display representative results for the heavy quark potential at $(L=32,\kappa_c=0.1)$ and $(L=32,\kappa_c=0.01)$ in \cref{fig:stringtension_DMRG}.

The string tension $\bsigma$ has dimensions of energy per unit length. In a theory with explicit Lorentz invariance, one may work in natural units $\hbar c = 1$, in which case $\bsigma$ has the same dimensions as a mass squared. In a lattice Hamiltonian formulation, however, one must study the dispersion relation of the elementary excitations in order to extract an equivalent dimensionless quantity $\zeta$. This allows us to define a physical string tension $\sigma = \zeta\,\bsigma$ with dimensions of energy squared, and hence to form the dimensionless ratio $\sqrt{\sigma}/m_1$, which is a universal quantity in the continuum \ac{QFT}.

Because the lattice theory possesses discrete translational invariance, when the system is placed in a finite box of length $L$, each eigenstate of the Hamiltonian ${\cal E}_i$ ($i=0,1,2,\dots$) can be assigned a quantized momentum
\begin{align}
p^{(n)} = \frac{2\pi n}{L}\,,
\end{align}
where the integer $n \in [-L/2, L/2]$. Energy eigenstates sharing the same momentum $p^{(n)}$ are labeled consecutively by an index $j=0,1,2,\dots$, so that ${\cal E}_i \equiv {\cal E}_j^{(n)}$. The state with $j=0$ and $n=0$ is identified as the vacuum with energy ${\cal E}_0$, while the higher $n=0$ levels help define the masses $m_j$ with $j=1,2,\dots$. Subtracting the vacuum energy, we define the renormalized energies
\begin{align}
E_j^{(n)} = {\cal E}_j^{(n)} - {\cal E}_0^{(0)}\,.
\end{align}
If the lattice theory describes a massive Lorentz-invariant \ac{QFT} near the critical point, then at sufficiently large volumes the low-energy spectrum is expected to satisfy the dispersion relation
\begin{align}
\bigl(E_j^{(n)}\bigr)^2
&= m_j^2 + \zeta^2 \bigl(p^{(n)}\bigr)^2
+ {\cal O}(n^4,L^{-4})\,,
\label{eq:edisp}
\end{align}
where the parameter $\zeta$ is independent of both $j$ and $n$. This same parameter was introduced above and serves to relate energy and length scales, playing the role of $\hbar c$ in the continuum dispersion relation.

\begin{figure}[t]
\centering
\includegraphics[width=0.47\textwidth]{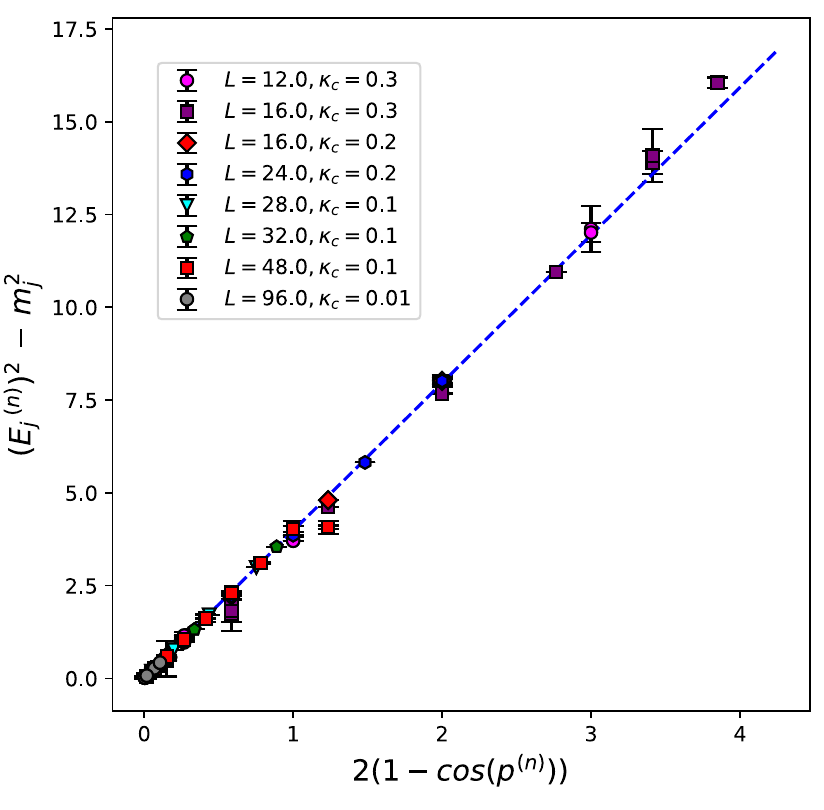}
\caption{Plot of the dispersion relation~\cref{eq:edisp}, demonstrating that all low-energy eigenvalues are consistent with a Lorentz-invariant form. The dashed line shows a linear fit to the $L=96$, $\kappa_c=0.01$ data; its slope is $\zeta^2$, yielding $\zeta = 1.995(7)$.}
\label{fig:edisp}
\end{figure}

We use \ac{DMRG} to compute the low-lying energy spectrum for various values of $L$ and $\kappa_c$, but we do not directly measure the momenta $p^{(n)}$. As a result, assigning the quantum numbers $j$ and $n$ to each energy level is nontrivial, particularly in the presence of numerical uncertainties. In our analysis, we first replace $(p^{(n)})^2$ in \cref{eq:edisp} by $2\bigl(1-\cos p^{(n)}\bigr)$, which is periodic over the Brillouin zone and coincides with the free-particle dispersion relation obtained from a naïve lattice discretization. We then label each nondegenerate state consecutively by an index $j$ and identify its energy as $m_j$. For each fixed $j$, successive values of $|n|$ are assigned to pairs of degenerate states whenever their energies are consistent with the dispersion relation in \cref{eq:edisp}. Degenerate states that do not satisfy this criterion are instead assigned to higher values of $j$ and $n$, as appropriate.

Consistency is checked by fitting $(E_j^{(n)})^2$ as a function of $2\bigl(1-\cos p^{(n)}\bigr)$. For small values of the momentum, a linear fit is sufficient; however, due to the presence of lattice artifacts and, in some cases, the high accuracy of the data, higher-order terms in the dispersion relation become resolvable. In these cases, a polynomial fit upto third order is required to obtain an acceptable description. The resulting assignments of $j$ and $n$ to the various energy levels are summarized in \cite{supp}. To demonstrate that all low-energy levels are consistent with \cref{eq:edisp} with a single parameter $\zeta$, we plot $(E_j^{(n)})^2 - m_j^2$ as a function of $2\bigl(1-\cos p^{(n)}\bigr)$ in \cref{fig:edisp}. The dashed line in this figure corresponds to a linear fit to the $L=96$, $\kappa_c=0.01$ data, yielding $\zeta = 1.995(7)$.

Our \ac{DMRG} analysis was performed using the ITensor software library in Julia \cite{ITensor,ITensor-r0.3}, and the fits were done using the NumPy~\cite{harris2020array} python package and the stats R package~\cite{Rcore}. For the \ac{DMRG}, we use cutoff parameters of $10^{-15}$ or smaller, perform 25 \ac{DMRG} sweeps for each data point, and allow a maximum bond dimension of 400. Varying the bond dimension between 300 and 400 or the cutoff between \(10^{-14}\) and \(10^{-15}\) changes the results by less than \(10^{-4}\).  Though this indicates that the \ac{DMRG} calculations are well converged, the resulting points deviate from the expected linear behavior in \cref{eq:stringT} due to various residual errors including the $\txtw$-dependence. These errors range from $0.002$ for $L=32$ to $0.0004$ for $L=96$ in the energy differences and are shown in \cref{fig:stringtension_DMRG}. Our results for ${\cal E}_0(\txtw)$, $m_1$, $m_2$, $\bsigma$ for several values of $L$ and $\kappa_c$ are listed in \Cref{tab:string_tension}. In this table, we also report the universal ratios $m_2/m_1$ and $\sqrt{\sigma}/m_1$. While $m_2/m_1$ approaches the value predicted by \ac{E8QFT}, we obtain the new universal dimensionless ratio $\sqrt{\sigma}/m_1 = 0.249(1)$ in the continuum \ac{QFT}.

\section{Asymptotic Freedom}
\label{sec5}

We can explore the physics of the massive continuum \ac{E8QFT} in the \ac{UV}. Since the critical lattice field theory flows to the \ac{IR} fixed point governed by the \ac{ICFT}, we expect the \ac{UV} physics of the massive continuum \ac{E8QFT} to be described by the same \ac{ICFT} (see \cref{fig:RGflow-trad}). If this expectation is correct, then, because the \ac{ICFT} is known to be the theory of free massless Majorana fermions, we expect these Majorana fermions to be a manifestation of the bosonic scalar adjoint `gauge bosons' attached to adjoint strings to project them onto the physical subspace of gauge-invariant states. We may then view the massive glueball states as arising through the confinement of these gauge bosons. In this interpretation, the $\SU(2)$ gauge theory on the plaquette chain also exhibits the phenomenon of asymptotic freedom. The viewpoint that the massive states arise as bound states of free Majorana fermions is well known (see for example \cite{Litvinov:2025geb}).

\begin{figure}[h]
\centering
\includegraphics[width=0.45\textwidth]{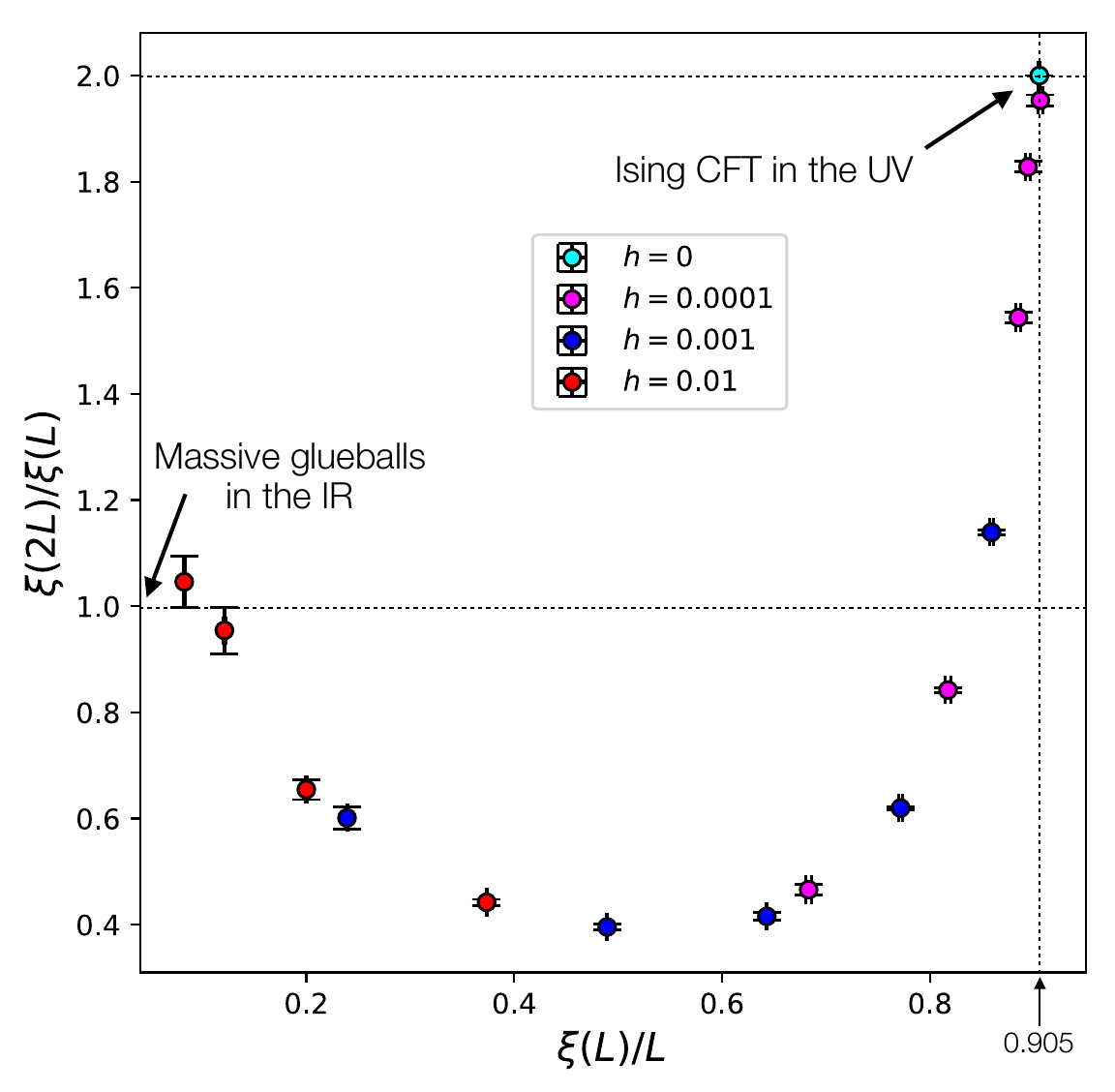}
\caption{Plot of the step-scaling function of the \ac{E8QFT} using its mapping to the two-dimensional Ising model in a uniform magnetic field, whose Euclidean lattice action is given in \cref{eq:IsingEuclid}. The correlation length $\xi(L)$ is extracted using Monte Carlo methods from the second-moment correlator of the subtracted field $(s_i - \bar{s})$, with $\bar{s}$ defined as the spin expectation value in the magnetic field. In the \ac{UV}, where we obtain the \ac{ICFT}, we obtain the universal ratio $\xi(L)/L = 0.905(1)$. Our result is consistent with the known value $0.9050488292(4)$, obtained by numerically evaluating the exact \ac{CFT} expressions for the torus two-point function and correlation-length amplitudes \cite{SalasSokal}.}
\label{fig:fssplot}
\end{figure}

Recently, it was discovered that asymptotic freedom and massive continuum \ac{QFT} may arise from exotic \ac{RG} flows, in which the \ac{UV} behavior is not governed by the \ac{IR} fixed point of the critical qubit-regularized lattice field theory, but rather by a crossover fixed point that the \ac{RG} trajectories approach before swinging away
\cite{Bhattacharya:2020gpm,Maiti:2023kpn}. To verify that no such exotic behavior occurs in the present case, using Monte Carlo methods, we have computed the step-scaling function of the \ac{E8QFT} using the Euclidean lattice formulation of the Ising model defined by the action
\begin{align}
S = -\beta_c \sum_{\langle i j\rangle} s_i s_j - h \sum_i s_i,
\label{eq:IsingEuclid}
\end{align}
where $\beta_c = \log(1+\sqrt{2})/2$ is the critical coupling at $h=0$. At criticality \cref{eq:IsingEuclid} describes the same physics as \cref{eq:TFIM} at long distances and low temperatures. The step-scaling function is defined in terms of the second-moment finite-size correlation length $\xi(L)$~\cite{PhysRevLett.75.1891}, computed from correlation functions of the subtracted Ising field $(s_i - \bar{s})$ to account for the nonzero average magnetization $\bar{s}$ when $h \neq 0$. The resulting step-scaling function is shown in \cref{fig:fssplot}.

\section{Conclusions}
\label{sec6}

In this work, we have argued that the qubit-regularized $\SU(2)$ lattice gauge theory (\ac{LGT}) on a plaquette chain possesses a continuum limit in which it describes a nontrivial massive \ac{QFT}. The massive excitations of this theory are one-dimensional analogs of glueballs, while in the \ac{UV} the theory becomes asymptotically free and is described by free massless Majorana fermions. These features arise through an exact mapping of the \ac{LGT} to the transverse-field Ising model (\ac{TFIM}). Importantly, although the Hamiltonian of our qubit-regularized \ac{LGT} differs from the conventional Kogut--Susskind Hamiltonian~\cite{Kogut:1974ag}, it nevertheless retains the two essential ingredients of lattice gauge theories: an electric-field term that induces confinement and a magnetic-field term that promotes deconfinement. In our construction, the former appears as a diagonal operator that assigns different energies to different representations, while the latter enters through an off-diagonal plaquette operator.

It is natural to ask whether analogous features extend to $\SU(3)$ \acp{LGT}. In this case, the qubit-regularized theory can be mapped to a class of three-state quantum clock models. Plaquette chains of this type have been explored recently in the context of Hamiltonian formulations and quantum computing~\cite{Ciavarella22, Ciavarella24}. However, the study of critical points from which massive relativistic \acp{QFT} may emerge through the \ac{RG} flow patterns illustrated in \cref{fig:RGflow-trad} remains largely unexplored. A straightforward $\SU(3)$ extension of the qubit-regularized lattice gauge theory proposed here can be tuned to the relativistic critical point of the two-dimensional three-state Potts model~\cite{RevModPhys.54.235}. Modifying the energies of the SU(3) electric fields along the chain then introduces a uniform Potts' magnetic field at criticality that favors one of the Potts states. This perturbation is relevant and can drive the theory to a massive relativistic \ac{QFT}~\cite{Delfino:2007tj,Lepori:2009ip}. We are currently exploring this massive \ac{QFT} using \ac{DMRG}.

Since qubit-regularized gauge theories are formulated within the Hamiltonian framework, they may realize both non-relativistic and relativistic quantum critical points~\cite{Orland:1989st,PhysRevB.98.205118}. An important direction for future work is to engineer qubit-regularized gauge theories in higher dimensions that exhibit relativistic quantum critical points. Relevant deformations of these theories could then lead to massive relativistic continuum quantum field theories analogous to those constructed here. 

\hbox{}\hfill\break

\acknowledgments

S.C. conceived of and carried out the initial investigations of this theory and created the outline of the manuscript. R.X.S. did the DMRG calculations. R.X.S. and S.C. did the Monte Carlo calculations, produced the figures, and produced the first draft of the paper.  S.C. and T.B. provided the interpretation of the model in terms of a pseudo-one-dimensional gauge theory. All three authors refined the manuscript and agree with its final version.

S.C. would like to thank Ribhu Kaul and Hansen Wu for bringing the physics of the massive E8 \ac{QFT} to his attention. He also thanks Peter Orland and Berndt M\"uller for helpful conversations about the manuscript. We acknowledge the use of AI assistance, specifically ChatGPT~\hbox{\cite{openai2025chatgpt}} and Gemini~\hbox{\cite{Google2026Gemini}}, in searching the literature relevant to SU(3) chains, refining the language and clarity of this manuscript, and providing citation to themselves, before our final rounds of manual review and revision by all authors. S.C. and R.X.S. are supported in part by the U.S. Department of Energy, Office of Science, Nuclear Physics program under Award No.\ DE-FG02-05ER41368. T.B. was supported by the U.S. Department of Energy, Office of Science, Office of High Energy Physics under contract number KA2401012 (LANLE83G) at Los Alamos National Laboratory operated by Triad National Security, LLC, for the National Nuclear Security Administration of the U.S. Department of Energy (Contract No.\ 89233218CNA000001).

\bibliographystyle{apsrev4-2} 
\showtitleinbib
\bibliography{fixapsbib,ref,refE8}

\clearpage
\onecolumngrid
\setcounter{page}{1}
\setcounter{table}{0}
\section*{Supplementary Material}

\begin{center}
{\bf Asymptotic-freedom and massive glueballs in a qubit-regularized SU(2) gauge theory}

\vskip0.1in

{\sl Rui Xian Siew, Shailesh Chandrasekharan, and Tanmoy Bhattacharya}

\end{center}

\begin{table*}[h]
\centering
\renewcommand{\arraystretch}{1.4}
\setlength{\tabcolsep}{4pt}
\begin{minipage}{0.3\textwidth}
\centering
\begin{tabular}{c|c|c|c}
\TopRule
${\cal E}_j^{(n)}/L$ & $\delta {\cal E}_j^{(n)}/L$ & $j$ & $n$  \\
\BotRule
\multicolumn{4}{c}{$L=12$, $\kappa_c=0.3$} \\
\MidRule
-1.52271e+00 & 5.6e-08 & 0 & 0 \\
-1.28658e+00 & 6.0e-08 & 1 & 0 \\
-1.27269e+00 & 8.2e-08 & 1 & 1 \\
-1.27269e+00 & 4.7e-08 & 1 & 1 \\
-1.23721e+00 & 7.7e-08 & 1 & 2 \\
-1.23721e+00 & 7.2e-08 & 1 & 2 \\
-1.19243e+00 & 7.4e-08 & 1 & 3 \\
-1.19243e+00 & 5.6e-08 & 1 & 3 \\
-1.14872e+00 & 5.7e-03 & 1 & 4 \\
-1.14964e+00 & 2.3e-03 & 1 & 4 \\
-1.14602e+00 & 2.8e-03 & 2 & 0 \\
-1.13551e+00 & 2.2e-05 & 2 & 1 \\
-1.13554e+00 & 8.5e-04 & 2 & 1 \\
\BotRule 
\end{tabular}
\end{minipage}
\begin{minipage}{0.3\textwidth}
\centering
\begin{tabular}{c|c|c|c}
\TopRule
${\cal E}_j^{(n)}/L$ & $\delta {\cal E}_j^{(n)}/L$ & $j$ & $n$  \\
\BotRule
\multicolumn{4}{c}{$L=16$, $\kappa_c=0.3$} \\
\MidRule
-1.52271e+00 & 1.7e-05 & 0 & 0 \\
-1.34561e+00 & 1.9e-06 & 1 & 0 \\
-1.33964e+00 & 1.8e-06 & 1 & 1 \\
-1.33964e+00 & 1.9e-06 & 1 & 1 \\
-1.32336e+00 & 1.8e-06 & 1 & 2 \\
-1.32336e+00 & 1.7e-06 & 1 & 2 \\
-1.30048e+00 & 1.8e-06 & 1 & 3 \\
-1.30048e+00 & 1.8e-06 & 1 & 3 \\
-1.27500e+00 & 1.8e-06 & 1 & 4 \\
-1.27500e+00 & 1.9e-06 & 1 & 4 \\
-1.25042e+00 & 1.5e-06 & 1 & 5 \\
-1.25042e+00 & 3.3e-05 & 1 & 5 \\
-1.23849e+00 & 5.2e-05 & 2 & 0 \\
-1.23519e+00 & 6.5e-05 & 2 & 1 \\
-1.23487e+00 & 3.2e-03 & 2 & 1 \\
-1.23008e+00 & 2.1e-03 & 1 & 6 \\
-1.22881e+00 & 4.7e-03 & 1 & 6 \\
-1.22671e+00 & 3.1e-03 & 2 & 2 \\
-1.22628e+00 & 1.9e-03 & 2 & 2 \\
-1.21602e+00 & 8.8e-04 & 1 & 7 \\
\BotRule
\end{tabular}
\end{minipage}
\begin{minipage}{0.3\textwidth}
\centering
\begin{tabular}{c|c|c|c}
\TopRule
${\cal E}_j^{(n)}/L$ & $\delta {\cal E}_j^{(n)}/L$ & $j$ & $n$  \\
\BotRule
\multicolumn{4}{c}{$L=16$, $\kappa_c=0.2$} \\
\MidRule
-1.43534e+00 & 1.2e-07 & 0 & 0 \\
-1.29242e+00 & 8.8e-08 & 1 & 0 \\
-1.28481e+00 & 4.9e-08 & 1 & 1 \\
-1.28481e+00 & 3.9e-08 & 1 & 1 \\
-1.26465e+00 & 1.1e-07 & 1 & 2 \\
-1.26465e+00 & 2.1e-08 & 1 & 2 \\
-1.23738e+00 & 5.2e-08 & 1 & 3 \\
-1.23738e+00 & 4.5e-08 & 1 & 3 \\
-1.20793e+00 & 4.4e-04 & 1 & 4 \\
-1.20783e+00 & 1.4e-03 & 1 & 4 \\
-1.20574e+00 & 7.8e-04 & 2 & 0 \\
-1.20067e+00 & 3.1e-04 & 2 & 1 \\
\BotRule
\end{tabular}
\end{minipage}
\caption{These tables show the assignment of the integers $j$ and $n$ to various energy levels. Note that the first two columns in each table give the energy density (not the energy) and its estimated error.}
\end{table*}

\begin{table*}[t]
\centering
\renewcommand{\arraystretch}{1.4}
\setlength{\tabcolsep}{4pt}
\begin{minipage}{0.3\textwidth}
\centering
\begin{tabular}{c|c|c|c}
\TopRule
${\cal E}_j^{(n)}/L$ & $\delta {\cal E}_j^{(n)}/L$ & $j$ & $n$  \\
\BotRule
\multicolumn{4}{c}{$L=24$, $\kappa_c=0.2$} \\
\MidRule
-1.43534e+00 & 4.4e-08 & 0 & 0 \\
-1.34006e+00 & 2.2e-06 & 1 & 0 \\
-1.33776e+00 & 1.1e-06 & 1 & 1 \\
-1.33776e+00 & 7.4e-08 & 1 & 1 \\
-1.33127e+00 & 3.5e-07 & 1 & 2 \\
-1.33127e+00 & 6.6e-07 & 1 & 2 \\
-1.32155e+00 & 1.3e-07 & 1 & 3 \\
-1.32155e+00 & 1.3e-07 & 1 & 3 \\
-1.30973e+00 & 1.0e-07 & 1 & 4 \\
-1.30973e+00 & 5.1e-08 & 1 & 4 \\
-1.29684e+00 & 6.6e-08 & 1 & 5 \\
-1.29684e+00 & 1.9e-07 & 1 & 5 \\
-1.28203e+00 & 1.1e-03 & 2 & 0 \\
-1.28370e+00 & 5.1e-04 & 1 & 6 \\
-1.28370e+00 & 5.0e-04 & 1 & 6 \\
-1.28065e+00 & 3.3e-04 & 2 & 1 \\
\BotRule
\end{tabular}
\end{minipage}
\begin{minipage}{0.3\textwidth}
\centering
\begin{tabular}{c|c|c|c}
\TopRule
${\cal E}_j^{(n)}/L$ & $\delta {\cal E}_j^{(n)}/L$ & $j$ & $n$  \\
\BotRule
\multicolumn{4}{c}{$L=28$, $\kappa_c=0.1$} \\
\MidRule
-1.35074e+00 & 1.6e-07 & 0 & 0 \\
-1.29421e+00 & 2.9e-06 & 1 & 0 \\
-1.29205e+00 & 1.3e-06 & 1 & 1 \\
-1.29205e+00 & 3.0e-07 & 1 & 1 \\
-1.28608e+00 & 7.7e-07 & 1 & 2 \\
-1.28608e+00 & 1.4e-07 & 1 & 2 \\
-1.27738e+00 & 7.4e-08 & 1 & 3 \\
-1.27738e+00 & 1.1e-07 & 1 & 3 \\
-1.26701e+00 & 1.2e-07 & 1 & 4 \\
-1.26701e+00 & 1.2e-07 & 1 & 4 \\
-1.25949e+00 & 2.3e-04 & 2 & 0 \\
-1.25817e+00 & 7.8e-05 & 2 & 1 \\
-1.25817e+00 & 1.0e-04 & 2 & 1 \\
\BotRule
\end{tabular}
\end{minipage}
\begin{minipage}{0.3\textwidth}
\centering
\begin{tabular}{c|c|c|c}
\TopRule
${\cal E}_j^{(n)}/L$ & $\delta {\cal E}_j^{(n)}/L$ & $j$ & $n$  \\
\BotRule
\multicolumn{4}{c}{$L=32$, $\kappa_c=0.1$} \\
\MidRule
-1.35074e+00 & 1.1e-07 & 0 & 0 \\
-1.30128e+00 & 1.8e-05 & 1 & 0 \\
-1.29983e+00 & 4.2e-06 & 1 & 1 \\
-1.29983e+00 & 1.0e-05 & 1 & 1 \\
-1.29574e+00 & 5.5e-06 & 1 & 2 \\
-1.29574e+00 & 2.9e-06 & 1 & 2 \\
-1.28962e+00 & 8.3e-07 & 1 & 3 \\
-1.28962e+00 & 7.6e-08 & 1 & 3 \\
-1.28214e+00 & 1.3e-07 & 1 & 4 \\
-1.28214e+00 & 7.1e-07 & 1 & 4 \\
-1.27386e+00 & 2.8e-05 & 1 & 5 \\
-1.27386e+00 & 1.2e-05 & 1 & 5 \\
-1.27089e+00 & 1.1e-04 & 2 & 0 \\
-1.27002e+00 & 5.5e-05 & 2 & 1 \\
\BotRule
\end{tabular}
\end{minipage}
\caption{These tables show the assignment of the integers $j$ and $n$ to various energy levels. Note that the first two columns in each table give the energy density (not the energy) and its estimated error.}
\end{table*}

\begin{table*}[b]
\centering
\renewcommand{\arraystretch}{1.4}
\setlength{\tabcolsep}{4pt}
\begin{minipage}{0.3\textwidth}
\centering
\begin{tabular}{c|c|c|c}
\TopRule
${\cal E}_j^{(n)}/L$ & $\delta {\cal E}_j^{(n)}/L$ & $j$ & $n$  \\
\BotRule
\multicolumn{4}{c}{$L=48$, $\kappa_c=0.1$} \\
\MidRule
-1.35074e+00 & 9.5e-07 & 0 & 0 \\
-1.31774e+00 & 2.3e-04 & 1 & 0 \\
-1.31735e+00 & 2.2e-04 & 1 & 1 \\
-1.31732e+00 & 4.6e-04 & 1 & 1 \\
-1.31607e+00 & 1.9e-04 & 1 & 2 \\
-1.31608e+00 & 2.4e-04 & 1 & 2 \\
-1.31407e+00 & 5.0e-05 & 1 & 3 \\
-1.31407e+00 & 1.3e-04 & 1 & 3 \\
-1.31148e+00 & 4.2e-05 & 1 & 4 \\
-1.31147e+00 & 3.8e-04 & 1 & 4 \\
-1.30841e+00 & 1.1e-04 & 1 & 5 \\
-1.30841e+00 & 5.2e-04 & 1 & 5 \\
-1.30501e+00 & 8.0e-05 & 1 & 6 \\
-1.30501e+00 & 2.3e-04 & 1 & 6 \\
-1.30137e+00 & 3.5e-05 & 1 & 7 \\
-1.30137e+00 & 1.2e-04 & 1 & 7 \\
-1.29747e+00 & 3.3e-04 & 1 & 8 \\
-1.29728e+00 & 1.8e-04 & 1 & 9 \\
-1.29729e+00 & 7.2e-04 & 1 & 9 \\
-1.29654e+00 & 9.6e-04 & 2 & 0 \\
-1.29749e+00 & 9.1e-04 & 1 & 8 \\
\BotRule
\end{tabular}
\end{minipage}
\begin{minipage}{0.3\textwidth}
\centering
\begin{tabular}{c|c|c|c}
\TopRule
${\cal E}_j^{(n)}/L$ & $\delta {\cal E}_j^{(n)}/L$ & $j$ & $n$  \\
\BotRule
\multicolumn{4}{c}{$L=96$, $\kappa_c=0.01$} \\
\MidRule
-1.27990e+00 & 4.5e-07 & 0 & 0 \\
-1.27505e+00 & 1.1e-04 & 1 & 0 \\
-1.27487e+00 & 4.3e-05 & 1 & 1 \\
-1.27488e+00 & 1.0e-04 & 1 & 1 \\
-1.27435e+00 & 3.3e-05 & 1 & 2 \\
-1.27435e+00 & 3.2e-05 & 1 & 2 \\
-1.27357e+00 & 1.3e-05 & 1 & 3 \\
-1.27357e+00 & 5.1e-06 & 1 & 3 \\
-1.27261e+00 & 1.5e-04 & 1 & 4 \\
-1.27261e+00 & 1.8e-04 & 1 & 4 \\
-1.27208e+00 & 3.2e-04 & 2 & 0 \\
-1.27196e+00 & 1.9e-04 & 2 & 1 \\
-1.27194e+00 & 3.1e-04 & 2 & 1 \\
-1.27161e+00 & 1.7e-04 & 2 & 2 \\
-1.27158e+00 & 3.4e-04 & 2 & 2 \\
-1.27160e+00 & 2.2e-04 & 1 & 5 \\
-1.27155e+00 & 3.5e-04 & 1 & 5 \\
-1.27108e+00 & 2.4e-04 & 3 & 0 \\
\BotRule
\end{tabular}
\end{minipage}
\caption{These tables show the assignment of the integers $j$ and $n$ to various energy levels. Note that the first two columns in each table give the energy density (not the energy) and its estimated error.}
\end{table*}

\end{document}